**Enhancement of the efficiency of acousto-optic Bragg diffraction due to optical activity. A case of $Pb_5Ge_3O_{11}$ crystals**

Oksana Mys, Myroslav Kostyrko, Ivan Orykhivskyi, Dmitro Adamenko, Ihor Skab and Rostyslav Vlokh

*O. G. Vlokh Institute of Physical Optics, 23 Dragomanov Street, 79005 Lviv, Ukraine*
Corresponding author: Rostyslav Vlokh

E-mail: vlokh@ifo.lviv.ua

**Abstract.** We show that the existence of optical activity in an optical material can lead to essential enhancement of acousto-optic (AO) figure of merit for this material. The reason is that the ellipticity of interacting optical eigenwaves approaches unity near the optic axis and so additional elasto-optic (EO) tensor components with relatively large values become involved into the effective EO coefficient. We demonstrate on the example of lead germanate crystals, $Pb_5Ge_3O_{11}$, that the increase in the efficiency of AO diffraction manifests itself for all the types of isotropic and anisotropic interactions, whenever the incident optical wave propagates close to the optic axis. We find that, in the particular case of diffraction in the interaction plane *XZ* of $Pb_5Ge_3O_{11}$ crystals, the maximal enhancement of the AO figure of merit takes place under conditions of the types V and VI of isotropic diffraction, with the AO figure of merit increasing from zero up to $13.3\times10^{-15}$ $s^3/kg$, and the type IX of anisotropic diffraction when the AO figure of merit increases more than twice (from $12.5\times10^{-15}$ up to $26.5\times10^{-15}$ $s^3/kg$). The maximal AO efficiency in the *XZ* interaction plane is reached at the types I and II of isotropic AO interactions. In these cases the AO figure of merit increases from $6.8\times10^{-15}$ up to $37.9\times10^{-15}$ $s^3/kg$ and from $31.1\times10^{-15}$ to $37.9\times10^{-15}$ $s^3/kg$, respectively.

## 1. Introduction

Acousto-optic (AO) diffraction is a well-known phenomenon [1, 2] that finds its applications in a number of optical technologies such as modulation of light [3], light beam deflection and scanning [3, 4], AO tunable filtering [5], etc. The AO diffraction consists in interaction of light with a phase grating of refractive index created in a material medium due to elasto-optic (EO) effect. The efficiency of AO diffraction is characterized by a so-called AO figure of merit [3],

$$M_2 = \frac{n_i^3 n_d^3 p_{eff}^2}{\rho v^3}, \quad (1)$$

which depends on the constitutive parameters of a material: the refractive indices of the incident ($n_i$) and diffracted ($n_d$) optical waves, the velocity $v$ of acoustic wave (AW), the effective EO coefficient $p_{eff}$, and the density $\rho$. In fact, in the case of low AO efficiency the AO figure of merit represents a proportionality coefficient between the intensity of diffracted optical waves and the AW power.

Usually, the effective EO coefficient is represented by cumbersome relations involving different EO tensor components. In spite of this fact, the effective EO has been derived for different point symmetry groups of crystals and different directions of propagation of interacting optical waves and AWs (see, e.g., Refs. [6, 7]). Nonetheless, we have shown recently that, in the case when optical eigenwaves are circularly polarized due to a presence of optical activity, the appropriate AO diffraction can be considered as a separate type of AO interactions between optical eigenwaves and AWs [8−10]. Moreover, it is a known fact that consideration of optical activity in one of the most efficient AO materials, $TeO_2$ crystal, gives rise to increase in the AO figure of merit from (600–800)×$10^{-15}$ $s^3$/kg for the linearly polarized optical waves up to 1200×$10^{-15}$ $s^3$/kg for the circularly polarized optical waves [11] whenever the incident and diffracted optical beams propagate in the vicinity of optic axis.

Paratellurite is not the only material where accounting for the optical activity increases the AO figure of merit. It has been found in the recent work [12] that the diffraction efficiency in optically active α-$HIO_3$ crystals also increases significantly when the incident and diffracted optical waves propagate along the directions close to the optic axis. Then the following questions appear: (i) Does a consideration of optical activity enhance the diffraction efficiency for all the types of isotropic and anisotropic AO interactions? and (ii) Is the increase in the AO figure of merit occurring for optical active crystals peculiar only for the $TeO_2$ and α-$HIO_3$ crystals – or this feature concerns the other optically active crystals, too? In our recent work [6], we have considered the anisotropy of AO figure of merit in the *XZ* plane of $Pb_5Ge_3O_{11}$ crystals without taking the optical activity into account. Using the approximation of linearly polarized optical eigenwaves, we have found that the maximal AO figure of merit for the isotropic diffraction is equal to 30.3×$10^{-15}$ $s^3$/kg. It is achieved for the AO diffraction of a so-called type II of extraordinary optical wave on the quasi-longitudinal AW. In the case of anisotropic AO diffraction, a corresponding maximum, 12.4×$10^{-15}$ $s^3$/kg, is reached for the AO interactions of a type IX with pure transverse AW, polarized perpendicular to *XZ* plane. In the present work we solve the problems mentioned above on the example of lead germanate crystals.

## 2. Methods of analysis

$Pb_5Ge_3O_{11}$ crystals undergo a proper ferroelectric phase transition at 450 K with the symmetry change $\bar{6} \Leftrightarrow 3$ [13]. The attenuation of AWs in these crystals at gigahertz frequencies is high enough [14], thus making lead germanate unsuitable for AO applications in the high-frequency range. However, the attenuation of AWs falls down to ~ 1 dB/cm in the megahertz range, thus leaving a possibility for wider applications of $Pb_5Ge_3O_{11}$ as a working material for various AO devices.

Lead germanate is optically active. The optical rotation occurring for the light propagation direction parallel to the optic axis is equal to $\pm 5.58$ deg/mm at $\lambda = 632.8$ nm [15]. The appropriate gyration tensor components are equal to $g_{33} = \pm 4.16\times10^{-5}$ [12] and $g_{11} = \pm 10.5\times10^{-5}$ at $\lambda = 632.8$ nm [16], and the signs of the both tensor components are the same. Switching of spontaneous electric polarization reverses the optical activity, since the ferroelectric domains are enantiomorphous. The ordinary and extraordinary refractive indices for the lead germanate crystals are equal to $n_o = 2.116$ and $n_e = 2.151$ at the optical wavelength 632.8 nm [12], and the density of this material amounts to $\rho = 7.33\times10^3$ kg/m$^3$ [17].

Dependences of the AW velocities on the angle $\theta$ between the *X* axis and the wavevector of AW have been obtained following from the Christoffel equation and using the elastic stiffnesses taken from Ref. [18]: $C_{11} = 68.4\times10^9$ N/m$^2$, $C_{12} = 26.8\times10^9$ N/m$^2$, $C_{13} = 17.9\times10^9$ N/m$^2$, $C_{33} = 94.3\times10^9$ N/m$^2$, $C_{14} = 0.0$, $C_{25} = 1.2\times10^9$ N/m$^2$, $C_{44} = 22.6$, and $C_{66} = 20.8\times10^9$ N/m$^2$. The effective EO coefficients have been calculated using the EO components determined in the earlier works: $p_{12} = 0.177\pm0.005$, $p_{13} = 0.119\pm0.006$, $p_{31} = 0.136\pm0.016$, $p_{14} = 0.005\pm0.006$, $p_{15} = 0.011\pm0.002$, $p_{16} = -0.008\pm0.001$, $p_{41} = 0.005\pm0.013$, $p_{51} = -0.002\pm0.013$, $p_{44} = -0.025\pm0.012$, $p_{45} = -0.003\pm0.014$ [19], and $p_{11} = 0.162\pm0.009$ and $p_{33} = 0.159\pm0.009$ [9].

$Pb_5Ge_3O_{11}$ belongs to the point symmetry group 3 under normal conditions. The EO tensor for this point symmetry group has a complicated form and contains 12 components. Since this crystal has the lowest symmetry among either optically uniaxial or isotropic crystalline materials, it can serve as a good model object for the studies of anisotropy of AO figure of merit. We will investigate this anisotropy only in the *XZ* plane in order to compare our results with the data of the earlier work [6] where the optical activity has not been taken into account. We remind that the axes of the coordinate system *XYZ* are parallel to the principal axes of optical indicatrix ellipsoid. They are related to the crystallographic coordinate system as $Z \,||\, [0001]$ and $X \,||\, [2\bar{1}\bar{1}0]$, while the *Y* axis is perpendicular to the *XZ* plane.

The method used for deriving the effective EO coefficient has been described in detail in our work [6]. The components of the strain tensor caused by the quasi-longitudinal (QL) AW are given by

$$e_1 = \cos\theta\cos\zeta_2,\, e_3 = \sin\theta\sin\zeta_2\,,\, e_5 = \sin(\zeta_2 + \theta)\,, \tag{2}$$

where $\zeta_2$ denotes the angle between the displacement vector and the *X* axis. The components of the strain tensor caused by one of quasi-transverse AWs polarized in the *XZ* plane and abbreviated as $QT_1$ read as

$$e_1 = -\cos\theta\sin\zeta_2,\, e_3 = \sin\theta\cos\zeta_2\,,\, e_5 = \cos(\zeta_2 + \theta)\;. \tag{3}$$

Finally, the strain components caused by a purely transverse AW $PT_2$ polarized parallel to the *Y* axis are as follows:

$$e_6 = \cos\theta e_6',\, e_4 = \sin\theta e_6'\,. \tag{4}$$

Here the relationships $e_6' = \dfrac{\partial u_L}{\partial Y} + \dfrac{\partial u_T}{\partial X'} = uK_{X'}\sin(\Omega t - K_{X'}X')$ and $u_T = u\cos(\Omega t - K_{X'}X')$ hold true, $X'$ is the axis parallel to the wavevector of AW in the coordinate system $X'YZ'$ rotated by the angle $\theta$ around the *Y* axis, $\Omega$ implies the AW frequency, $K_{X'}$ the AW vector, $u_L$ and $u_T$ are respectively the longitudinal and transverse components of the unit displacement vector $u$, and *t* denotes the time.

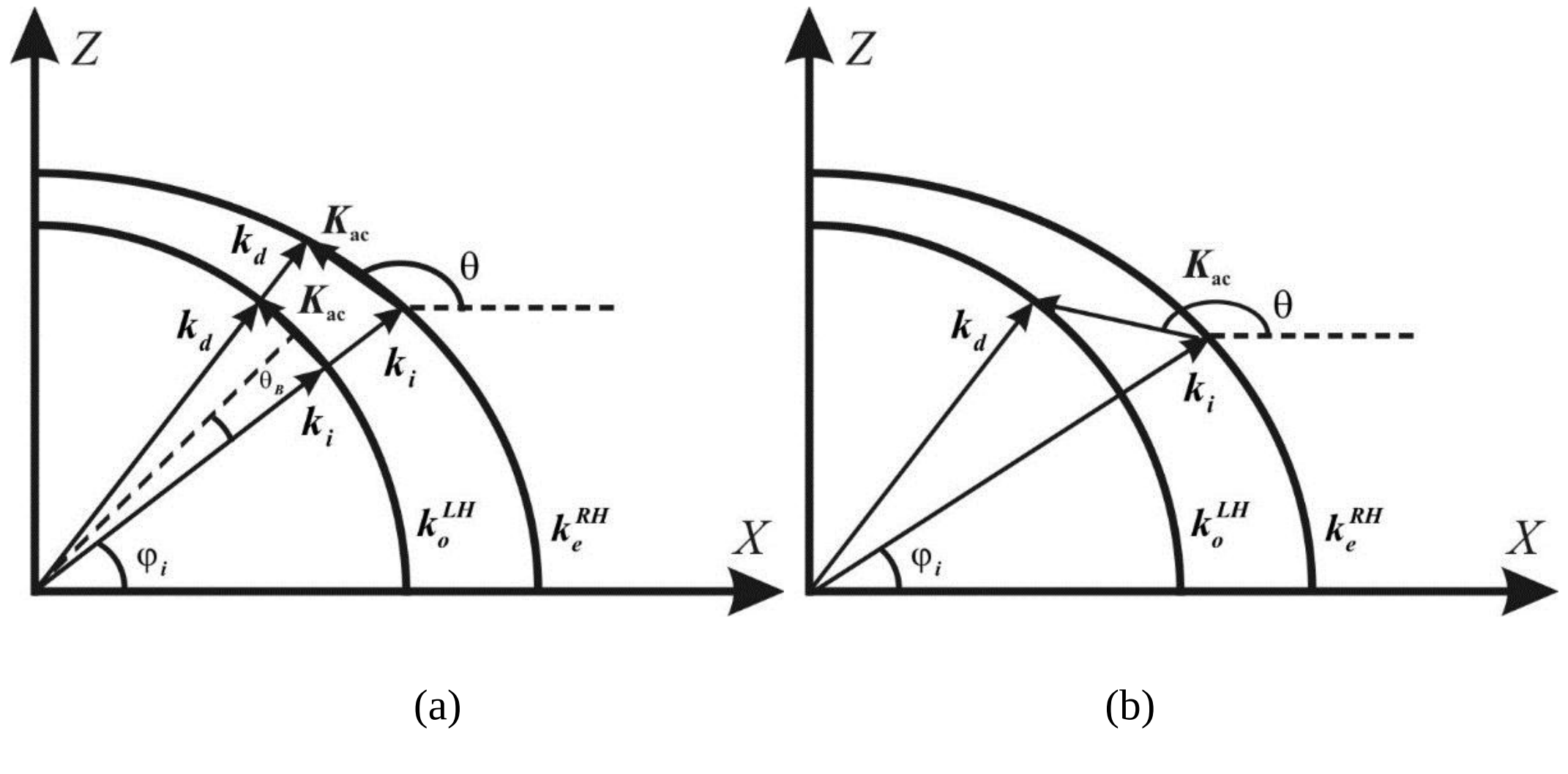

(a) (b)

Figure 1. Schematic representation of phase matching conditions for the isotropic (a) and anisotropic (b) AO diffractions in $Pb_5Ge_3O_{11}$: $k_i$, $k_d$ and $K_{ac}$ are wavevectors of respectively incident and diffracted optical waves and AW, $k_o^{LH}$ and $k_e^{RH}$ are cross sections of the indicative surfaces of wavevectors for left-handed ordinary wave and right-handed extraordinary wave, respectively.

The phase matching conditions for the isotropic and anisotropic diffractions are illustrated in Figure 1. Without accounting for the optical activity, the effective EO coefficients are as follows.

(I) The type I of isotropic AO interaction of the quasi-longitudinal (QL) AW with the optical wave polarized parallel to the *Y* axis:

$$p_{eff}^{2(I)} = 0.5\left(p_{12}\cos\theta\cos\zeta_2 + p_{13}\sin\theta\sin\zeta_2 + p_{25}\sin(\zeta_2+\theta)\right)^2, \quad (5)$$

(II) The type II of isotropic interaction of the QL AW with the optical wave polarized in the *XZ* plane:

$$\begin{aligned} p_{eff}^{2\,(II)} &= 0.5\cos^2(\theta_B+\theta)\begin{bmatrix}(p_{11}\cos\theta\cos\zeta_2 + p_{13}\sin\theta\sin\zeta_2 - p_{25}e_5)^2\cos^2\theta \\ +(p_{44}\sin(\zeta_2+\theta) - p_{52}\cos\theta\cos\zeta_2)^2\sin^2\theta\end{bmatrix} \\ &+0.5(p_{44}\sin(\zeta_2+\theta) - p_{52}\cos\theta\cos\zeta_2)\sin 2\theta \\ &\times\begin{bmatrix}(p_{11}\cos\theta\cos\zeta_2 + p_{13}\sin\theta\sin\zeta_2 - p_{25}\sin(\zeta_2+\theta))\cos^2\theta \\ +(p_{31}\cos\theta\cos\zeta_2 + p_{33}\sin\theta\sin\zeta_2)^2\sin^2\theta\end{bmatrix} \\ &+0.5\sin^2(\theta_B+\theta)\begin{bmatrix}(p_{44}\sin(\zeta_2+\theta) - p_{52}\cos\theta\cos\zeta_2)^2\cos^2\theta \\ +(p_{31}\cos\theta\cos\zeta_2 + p_{33}\sin\theta\sin\zeta_2)^2\sin^2\theta\end{bmatrix} \end{aligned}, \quad (6)$$

where $\theta_B$ is the Bragg angle (taken to be equal to 0.5 deg) and $\varphi_i$ denotes the angle between the *X* axis and the propagation direction of the incident optical wave;

(III) The type III of isotropic interaction of the quasi-transverse AW $QT_1$ polarized in the *XZ* plane with the optical wave polarized parallel to the *Y* axis:

$$p_{eff}^{2(III)} = 0.5\left(p_{13}\sin\theta\cos\zeta_2 - p_{12}\cos\theta\sin\zeta_2 + p_{25}\cos(\zeta_2+\theta)\right)^2; \quad (7)$$

(IV) The type IV of isotropic interaction of the AW $QT_1$ polarized in the *XZ* plane with the optical wave polarized in the *XZ* plane:

$$p_{eff}^{2(IV)} = 0.5\cos^2(\theta_B+\theta)\begin{bmatrix}(p_{13}\sin\theta\cos\zeta_2 - p_{11}\cos\theta\sin\zeta_2 \\ -p_{25}\cos(\zeta_2+\theta))^2\cos^2\theta \\ +(p_{52}\cos\theta\sin\zeta_2 + p_{44}\cos(\zeta_2+\theta))^2\sin^2\theta\end{bmatrix}$$

$$+0.5(p_{52}\cos\theta\sin\zeta_2 + p_{44}\cos(\zeta_2+\theta))$$

$$\times\sin 2\theta\begin{bmatrix}(p_{13}\sin\theta\cos\zeta_2 - p_{11}\cos\theta\sin\zeta_2 \\ -p_{25}\cos(\zeta_2+\theta))\cos^2\theta \\ +(p_{33}\sin\theta\cos\zeta_2 - p_{31}\cos\theta\sin\zeta_2)^2\sin^2\theta\end{bmatrix} \quad ; (8)$$

$$+0.5\sin^2(\theta_B+\theta)\begin{bmatrix}(p_{52}\cos\theta\sin\zeta_2 + p_{44}\cos(\zeta_2+\theta))^2\cos^2\theta \\ +(p_{33}\sin\theta\cos\zeta_2 - p_{31}\cos\theta\sin\zeta_2)^2\sin^2\theta\end{bmatrix}$$

(V) The type V of isotropic interaction of the AW $PT_2$ polarized parallel to the *Y* axis with the optical wave polarized parallel to the *Y* axis:

$$p_{eff}^{2\,(V)} = 0.5(p_{26}e_6 + p_{24}e_4)^2; \quad (9)$$

(VI) The type VI of isotropic interaction of the AW $PT_2$ polarized parallel to the *Y* axis with the optical wave polarized in the *XZ* plane:

$$p_{eff}^{2(VI)} = 0.5\cos^2(\theta_B+\theta)\begin{bmatrix}(p_{14}\sin\theta + p_{16}\cos\theta)^2\cos^2\theta \\ +(p_{41}\cos\theta - p_{45}\sin\theta)^2\sin^2\theta\end{bmatrix}$$

$$+0.5(p_{41}\cos\theta - p_{45}\sin\theta)(p_{14}\sin\theta + p_{16}\cos\theta)\cos^2\theta\sin 2\theta\,; (10)$$

$$+0.5(p_{41}\cos\theta - p_{45}\sin\theta)^2\cos^2\theta\sin^2(\theta_B+\theta)$$

(VII) The type VII of anisotropic interaction with the QL AW (where the polarization of incident optical wave belongs to the *XZ* plane and the polarization of diffracted wave is parallel to the *Y* axis):

$$p_{eff}^{2(VII)} = 0.5\begin{Bmatrix}\left(p_{14}\sin(\zeta_2+\theta) - p_{16}\cos\theta\cos\zeta_2\right)^2\cos^2\theta \\ +\left(p_{41}\cos\theta\cos\zeta_2 + p_{45}\sin(\zeta_2+\theta)\right)^2\sin^2\theta \\ +\left(p_{14}\sin(\zeta_2+\theta) - p_{16}\cos\theta\cos\zeta_2\right) \\ \times\left(p_{41}\cos\theta\cos\zeta_2 + p_{45}\sin(\zeta_2+\theta)\right)\sin 2\theta\end{Bmatrix}; \quad (11)$$

(VIII) The type VIII of anisotropic interaction with the AW $QT_1$:

$$p_{eff}^{2(VIII)} = 0.5\left\{\begin{array}{l}\left(p_{16}\cos\theta\sin\zeta_2 + p_{14}\cos(\zeta_2+\theta)\right)^2\cos^2\theta \\ +\left(p_{45}\cos(\zeta_2+\theta) - p_{41}\cos\theta\sin\zeta_2\right)^2\sin^2\theta \\ +\left(p_{16}\cos\theta\sin\zeta_2 + p_{14}e_5\right)\cos(\zeta_2+\theta) \\ \times\left(p_{45}\cos(\zeta_2+\theta) - p_{41}\cos\theta\sin\zeta_2\right)\sin 2\theta\end{array}\right\}; \quad (12)$$

(IX) The type IX of anisotropic interaction with the AW PT$_2$:

$$p_{eff}^{2(IX)} = 0.5\left\{\begin{array}{l}\left(p_{14}\sin\theta + p_{16}\cos\theta\right)^2\cos^2\theta \\ +\left(p_{44}\sin\theta + p_{52}\cos\theta\right)^2\sin^2\theta \\ +\left(p_{25}\sin\theta + p_{66}\cos\theta\right) \\ \times\left(p_{44}\sin\theta + p_{52}\cos\theta\right)\sin 2\theta\end{array}\right\}. \quad (13)$$

Notice that the appropriate relations obtained in our recent work [9] contain an error. Namely, the projections of the electric field of diffracted wave and the induction of incident optical wave have not been taken into account there. Nevertheless, this error has not affected notably the main results of Ref. [9].

We have considered the optical activity with taking a nonzero ellipticity of eigenwaves into account. The ellipticity is equal to $\chi = \frac{1}{2G}\left((n_o^2 - n'^2_e) - \sqrt{(n_o^2 - n'^2_e) + 4G^2}\right)$ [20], where $G$ is the scalar gyration parameter. Both the parameter $G = g_{33}\sin^2\varphi_i + g_{11}\cos^2\varphi_i$ and the refractive index $n'^2_e = \frac{n_o^2 n_e^2}{n_e^2\cos^2\varphi_i + n_o^2\sin^2\varphi_i}$ are dependent on the propagation direction of the incident optical wave, where $\varphi_i$ is the angle between the $X$ axis and the propagation direction of the incident optical wave. Since the angle between the incident and diffracted optical waves is small enough (~ 1 deg), one can neglect it. Accounting for the ellipticity of eigenwaves for the type I of AO interaction leads to appearance of three components of the electric field of diffracted optical wave, instead of a single component $E_2$ as with the optical activity neglected:

$$\left\{\begin{array}{l}E_1 = (\chi\Delta B_{11}D_2\cos\theta + \chi\Delta B_{13}D_2\sin\theta)\cos\theta \\ E_2 = \chi\Delta B_{12}D_2\cos\theta + \Delta B_{22}D_2 + \chi\Delta B_{23}D_2\sin\theta \\ E_3 = (\chi\Delta B_{13}D_2\cos\theta + \chi\Delta B_{33}D_2\sin\theta)\sin\theta\end{array}\right. . \quad (14)$$

Here we have

$$\Delta B_{11} = p_{11}e_1 + p_{13}e_3 + p_{15}e_5,\ \Delta B_{12} = -p_{16}e_6 + p_{14}e_5,$$
$$\Delta B_{13} = p_{51}e_1 + p_{55}e_5,\ \Delta B_{23} = p_{41}e_1 + p_{45}e_5, \tag{15}$$
$$\Delta B_{22} = p_{21}e_1 + p_{23}e_3 + p_{25}e_5,\ \Delta B_{33} = p_{31}e_1 + p_{33}e_3.$$

The effective EO coefficient can be extracted from the relation $E^2 = E_1^2 + E_2^2 + E_3^2$.

The same is true for the other types of AO interaction. Then for the type I of interaction we have

$$p_{eff}^{2(I)} = 0.5\left(p_{12}\cos\theta\cos\zeta_2 + p_{13}\sin\theta\sin\zeta_2 + p_{25}\sin(\zeta_2+\theta)\right)^2$$
$$+0.5\chi^2\begin{pmatrix} \left(p_{11}\cos\theta\cos\zeta_2 + p_{13}\sin\theta\sin\zeta_2 - p_{25}\sin(\zeta_2+\theta)\right)^2\cos^4\theta \\ +\left(p_{31}\cos\theta\cos\zeta_2 + p_{33}\sin\theta\sin\zeta_2\right)^2\sin^4\theta \\ +0.5\left(p_{44}\sin(\zeta_2+\theta) - p_{52}\cos\theta\cos\zeta_2\right)\sin^2 2\theta \\ +\left(p_{11}\cos\theta\cos\zeta_2 + p_{13}\sin\theta\sin\zeta_2 - p_{25}\sin(\zeta_2+\theta)\right) \\ \times\left(p_{44}\sin(\zeta_2+\theta) - p_{52}\cos\theta\cos\zeta_2\right)\sin 2\theta\cos^2\theta \\ +\left(p_{44}\sin(\zeta_2+\theta) - p_{52}\cos\theta\cos\zeta_2\right) \\ \times\left(p_{31}\cos\theta\cos\zeta_2 + p_{33}\sin\theta\sin\zeta_2\right)\sin 2\theta\sin^2\theta \\ +\left(p_{14}\sin(\zeta_2+\theta) - p_{16}\cos\theta\cos\zeta_2\right)^2\cos^2\theta \\ +\left(p_{41}\cos\theta\cos\zeta_2 + p_{45}\sin(\zeta_2+\theta)\right)^2\sin^2\theta \\ +\left(p_{14}\sin(\zeta_2+\theta) - p_{16}\cos\theta\cos\zeta_2\right) \\ \times\left(p_{41}\cos\theta\cos\zeta_2 + p_{45}\sin(\zeta_2+\theta)\right)\sin 2\theta \end{pmatrix}$$
$$+\chi\left(p_{12}\cos\theta\cos\zeta_2 + p_{13}\sin\theta\sin\zeta_2 + p_{25}\sin(\zeta_2+\theta)\right)$$
$$\times\begin{pmatrix} \left(p_{14}\sin(\zeta_2+\theta) - p_{16}\cos\theta\cos\zeta_2\right)\cos\theta \\ +\left(p_{41}\cos\theta\cos\zeta_2 + p_{45}\sin(\zeta_2+\theta)\right)\sin\theta \end{pmatrix} .\tag{16}$$

The effective EO coefficient for the type II of AO interaction can be written as

$$p_{eff}^{2(II)} = 0.5\cos^2(\theta_B+\theta)\begin{bmatrix}(p_{11}\cos\theta\cos\zeta_2 + p_{13}\sin\theta\sin\zeta_2 \\ -p_{25}\sin(\zeta_2+\theta))^2\cos^2\theta \\ +(p_{44}\sin(\zeta_2+\theta) \\ -p_{52}\cos\theta\cos\zeta_2)^2\sin^2\theta\end{bmatrix}$$

$$+0.5(p_{44}\sin(\zeta_2+\theta) - p_{52}\cos\theta\cos\zeta_2)\sin 2\theta$$

$$\times\begin{bmatrix}(p_{11}\cos\theta\cos\zeta_2 + p_{13}\sin\theta\sin\zeta_2 - p_{25}\sin(\zeta_2+\theta))\cos^2\theta \\ +(p_{31}\cos\theta\cos\zeta_2 + p_{33}\sin\theta\sin\zeta_2)\sin^2\theta\end{bmatrix}$$

$$+0.5\sin^2(\theta_B+\theta)\begin{bmatrix}(p_{44}\sin(\zeta_2+\theta) - p_{52}\cos\theta\cos\zeta_2)^2\cos^2\theta \\ +(p_{31}\cos\theta\cos\zeta_2 + p_{33}\sin\theta\sin\zeta_2)^2\sin^2\theta\end{bmatrix}$$

$$+\chi\left\{\begin{matrix}\cos(\theta_B+\theta)\begin{bmatrix}(p_{11}\cos\theta\cos\zeta_2 + p_{13}\sin\theta\sin\zeta_2 - p_{25}\sin(\zeta_2+\theta)) \\ \times(p_{14}\sin(\zeta_2+\theta) - p_{16}\cos\theta\cos\zeta_2)\cos^2\theta \\ +(p_{44}\sin(\zeta_2+\theta) - p_{52}\cos\theta\cos\zeta_2) \\ \times(p_{41}\cos\theta\cos\zeta_2 + p_{45}\sin(\zeta_2+\theta))\sin^2\theta\end{bmatrix} \\ +\sin(\theta_B+\theta)\begin{bmatrix}(p_{44}\sin(\zeta_2+\theta) - p_{52}\cos\theta\cos\zeta_2) \\ \times(p_{14}\sin(\zeta_2+\theta) - p_{16}\cos\theta\cos\zeta_2)\cos^2\theta \\ +(p_{31}\cos\theta\cos\zeta_2 + p_{33}\sin\theta\sin\zeta_2) \\ \times(p_{41}\cos\theta\cos\zeta_2 + p_{45}\sin(\zeta_2+\theta))\sin^2\theta\end{bmatrix}\end{matrix}\right\} .(17)$$

$$+0.5\chi^2\begin{bmatrix}(p_{14}\sin(\zeta_2+\theta) - p_{16}\cos\theta\cos\zeta_2)^2\cos^2\theta \\ +(p_{41}\cos\theta\cos\zeta_2 + p_{45}\sin(\zeta_2+\theta))^2\sin^2\theta\end{bmatrix}$$

The type III of interaction is determined by the effective EO coefficient

$$
\begin{aligned}
p_{eff}^{2(III)} &= 0.5\left(p_{13}\sin\theta\cos\zeta_2 - p_{12}\cos\theta\sin\zeta_2 + p_{25}\cos(\zeta_2+\theta)\right)^2 \\
&+0.5\chi^2\left(\begin{array}{l}
\left(p_{13}\sin\theta\cos\zeta_2 - p_{11}\cos\theta\sin\zeta_2 - p_{25}\cos(\zeta_2+\theta)\right)^2\cos^4\theta \\
+\left(p_{33}\sin\theta\cos\zeta_2 - p_{31}\cos\theta\sin\zeta_2\right)^2\sin^4\theta \\
+0.5\left(p_{52}\cos\theta\sin\zeta_2 + p_{44}\cos(\zeta_2+\theta)\right)\sin^2 2\theta \\
+\left(p_{13}\sin\theta\cos\zeta_2 - p_{11}\cos\theta\sin\zeta_2 - p_{25}\cos(\zeta_2+\theta)\right) \\
\times\left(p_{52}\cos\theta\sin\zeta_2 + p_{44}\cos(\zeta_2+\theta)\right)\sin 2\theta\cos^2\theta \\
+\left(p_{52}\cos\theta\sin\zeta_2 + p_{44}\cos(\zeta_2+\theta)\right) \\
\times\left(p_{33}\sin\theta\cos\zeta_2 - p_{31}\cos\theta\sin\zeta_2\right)\sin 2\theta\sin^2\theta \\
+\left(p_{16}\cos\theta\sin\zeta_2 + p_{14}\cos(\zeta_2+\theta)\right)^2\cos^2\theta \\
+\left(p_{45}\cos(\zeta_2+\theta) - p_{41}\cos\theta\sin\zeta_2\right)^2\sin^2\theta \\
+\left(p_{16}\cos\theta\sin\zeta_2 + p_{14}\cos(\zeta_2+\theta)\right) \\
\times\left(p_{45}\cos(\zeta_2+\theta) - p_{41}\cos\theta\sin\zeta_2\right)\sin 2\theta
\end{array}\right) \\
&+\chi\left(p_{13}\sin\theta\cos\zeta_2 - p_{12}\cos\theta\sin\zeta_2 + p_{25}\cos(\zeta_2+\theta)\right) \\
&\times\left(\left(p_{16}\cos\theta\sin\zeta_2 + p_{14}\cos(\zeta_2+\theta)\right)\cos\theta + \left(\begin{array}{l} p_{45}\cos(\zeta_2+\theta) \\ -p_{41}\cos\theta\sin\zeta_2 \end{array}\right)\sin\theta\right). \quad (18)
\end{aligned}
$$

For the type IV of AO interaction the effective EO coefficient reads as

$$p_{eff}^{2(IV)} = 0.5\cos^2(\theta_B+\theta)\begin{bmatrix}(p_{13}\sin\theta\cos\zeta_2 - p_{11}\cos\theta\sin\zeta_2 \\ -p_{25}\cos(\zeta_2+\theta))^2\cos^2\theta \\ +(p_{52}\cos\theta\sin\zeta_2 + p_{44}\cos(\zeta_2+\theta))^2\sin^2\theta\end{bmatrix}$$

$$+0.5(p_{52}\cos\theta\sin\zeta_2 + p_{44}\cos(\zeta_2+\theta))$$

$$\times\sin 2\theta\begin{bmatrix}(p_{13}\sin\theta\cos\zeta_2 - p_{11}\cos\theta\sin\zeta_2 - p_{25}\cos(\zeta_2+\theta))\cos^2\theta \\ +(p_{33}\sin\theta\cos\zeta_2 - p_{31}\cos\theta\sin\zeta_2)^2\sin^2\theta\end{bmatrix}$$

$$+0.5\sin^2(\theta_B+\theta)\begin{bmatrix}(p_{52}\cos\theta\sin\zeta_2 + p_{44}\cos(\zeta_2+\theta))^2\cos^2\theta \\ +(p_{33}\sin\theta\cos\zeta_2 - p_{31}\cos\theta\sin\zeta_2)^2\sin^2\theta\end{bmatrix}$$

$$\chi\begin{Bmatrix}\cos(\theta_B+\theta)\begin{bmatrix}(p_{13}\sin\theta\cos\zeta_2 - p_{11}\cos\theta\sin\zeta_2 - p_{25}\cos(\zeta_2+\theta)) \\ \times(p_{16}\cos\theta\sin\zeta_2 + p_{14}\cos(\zeta_2+\theta))\cos^2\theta \\ +(p_{52}\cos\theta\sin\zeta_2 + p_{44}\cos(\zeta_2+\theta)) \\ \times(p_{45}\cos(\zeta_2+\theta) - p_{41}\cos\theta\sin\zeta_2)\sin^2\theta\end{bmatrix} \\ +\sin(\theta_B+\theta)\begin{bmatrix}(p_{52}\cos\theta\sin\zeta_2 + p_{44}\cos(\zeta_2+\theta)) \\ \times(p_{16}\cos\theta\sin\zeta_2 + p_{14}\cos(\zeta_2+\theta))\cos^2\theta \\ +(p_{33}\sin\theta\cos\zeta_2 - p_{31}\cos\theta\sin\zeta_2) \\ \times(p_{45}\cos(\zeta_2+\theta) - p_{41}\cos\theta\sin\zeta_2)\sin^2\theta\end{bmatrix}\end{Bmatrix} . \quad (19)$$

$$+0.5\chi^2\begin{bmatrix}(p_{16}\cos\theta\sin\zeta_2 + p_{14}\cos(\zeta_2+\theta))^2\cos^2\theta \\ +(p_{45}\cos(\zeta_2+\theta) - p_{41}\cos\theta\sin\zeta_2)^2\sin^2\theta\end{bmatrix}$$

The type V of AO interaction is determined by the effective EO coefficient

$$p_{eff}^{2(V)} = 0.5\left(p_{14}\sin\theta + p_{16}\cos\theta\right)^2$$
$$+0.5\chi^2\begin{pmatrix}\left(p_{14}\sin\theta + p_{16}\cos\theta\right)^2\cos^4\theta \\ +0.5\left(p_{41}\cos\theta - p_{45}\sin\theta\right)^2\sin^2 2\theta \\ +\left(p_{14}\sin\theta + p_{16}\cos\theta\right) \\ \times\left(p_{41}\cos\theta - p_{45}\sin\theta\right)\sin 2\theta\cos^2\theta \\ +\left(p_{25}\sin\theta + p_{66}\cos\theta\right)^2\cos^2\theta \\ +\left(p_{44}\sin\theta + p_{52}\cos\theta\right)^2\sin^2\theta \\ +\left(p_{25}\sin\theta + p_{66}\cos\theta\right) \\ \times\left(p_{44}\sin\theta + p_{52}\cos\theta\right)\sin 2\theta\end{pmatrix}$$
$$-\chi\left(p_{14}\sin\theta + p_{16}\cos\theta\right)\begin{pmatrix}\left(p_{25}\sin\theta + p_{66}\cos\theta\right)\cos\theta \\ +\left(p_{44}\sin\theta + p_{52}\cos\theta\right)\sin\theta\end{pmatrix}. \quad (20)$$

Finally, the effective EO coefficient for the type VI of isotropic AO interaction can be written as

$$p_{eff}^{2(VI)} = 0.5\cos^2(\theta_B + \theta)\begin{bmatrix}(p_{14}\sin\theta + p_{16}\cos\theta)^2\cos^2\theta \\ +(p_{41}\cos\theta - p_{45}\sin\theta)^2\sin^2\theta\end{bmatrix}$$
$$+0.5(p_{41}\cos\theta - p_{45}\sin\theta)(p_{14}\sin\theta + p_{16}\cos\theta)\cos^2\theta\sin 2(\theta_B + \theta)$$
$$+0.5(p_{41}\cos\theta - p_{45}\sin\theta)^2\cos^2\theta\sin^2(\theta_B + \theta) \quad .(21)$$
$$+\chi\begin{Bmatrix}\cos(\theta_B + \theta)\begin{bmatrix}(p_{14}\sin\theta + p_{16}\cos\theta)(p_{25}\sin\theta + p_{66}\cos\theta)\cos^2\theta \\ +(p_{41}\cos\theta - p_{45}\sin\theta)(p_{44}\sin\theta + p_{52}\cos\theta)\sin^2\theta\end{bmatrix} \\ +(p_{41}\cos\theta - p_{45}\sin\theta)(p_{25}\sin\theta + p_{66}\cos\theta)\cos^2\theta\sin(\theta_B + \theta)\end{Bmatrix}$$
$$+0.5\chi^2\begin{bmatrix}(p_{25}\sin\theta + p_{66}\cos\theta)^2\cos^2\theta + (p_{44}\sin\theta + p_{52}\cos\theta)^2\sin^2\theta \\ +(p_{14}\sin\theta + p_{16}\cos\theta)^2\end{bmatrix}$$

The effective EO coefficients for the anisotropic types VII–IX of AO interaction are as follows:

$$p_{eff}^{2(VII)} = 0.5\left\{\begin{array}{l}\left(p_{14}\sin(\zeta_2+\theta)-p_{16}\cos\theta\cos\zeta_2\right)^2\cos^2\varphi_i \\ +\left(p_{41}\cos\theta\cos\zeta_2+p_{45}\sin(\zeta_2+\theta)\right)^2\sin^2\varphi_i \\ +\left(p_{14}\sin(\zeta_2+\theta)-p_{16}\cos\theta\cos\zeta_2\right) \\ \times\left(p_{41}\cos\theta\cos\zeta_2+p_{45}\sin(\zeta_2+\theta)\right)\sin 2\varphi_i \\ +\chi^2\left(\begin{array}{l}\left(p_{14}\sin(\zeta_2+\theta)-p_{16}\cos\theta\cos\zeta_2\right)^2\cos^2\varphi_i \\ +\left(p_{41}\cos\theta\cos\zeta_2+p_{45}\sin(\zeta_2+\theta)\right)^2\sin^2\varphi_i\end{array}\right)\end{array}\right\}, \quad (22)$$

$$p_{eff}^{2(VIII)} = 0.5\left\{\begin{array}{l}\left(p_{16}\cos\theta\sin\zeta_2+p_{14}\cos(\zeta_2+\theta)\right)^2\cos^2\varphi_i \\ +\left(p_{45}\cos(\zeta_2+\theta)-p_{41}\cos\theta\sin\zeta_2\right)^2\sin^2\varphi_i \\ +\left(p_{16}\cos\theta\sin\zeta_2+p_{14}e_5\right)\cos(\zeta_2+\theta) \\ \times\left(p_{45}\cos(\zeta_2+\theta)-p_{41}\cos\theta\sin\zeta_2\right)\sin 2\varphi_i \\ +\chi^2\left(\begin{array}{l}\left(p_{16}\cos\theta\sin\zeta_2+p_{14}\cos(\zeta_2+\theta)\right)^2\cos^2\varphi_i \\ +\left(p_{45}\cos(\zeta_2+\theta)-p_{41}\cos\theta\sin\zeta_2\right)^2\sin^2\varphi_i\end{array}\right)\end{array}\right\}, \quad (23)$$

and

$$p_{eff}^{2(IX)} = 0.5\left\{\begin{array}{l}\left(p_{25}\sin\theta+p_{66}\cos\theta\right)^2\cos^2\varphi_i \\ +\left(p_{44}\sin\theta+p_{52}\cos\theta\right)^2\sin^2\varphi_i \\ +\left(p_{25}\sin\theta+p_{66}\cos\theta\right)\left(p_{44}\sin\theta+p_{52}\cos\theta\right)\sin 2\varphi_i \\ +\chi^2\left(\begin{array}{l}\left(p_{25}\sin\theta+p_{66}\cos\theta\right)^2\cos^2\varphi_i \\ +\left(p_{44}\sin\theta+p_{52}\cos\theta\right)^2\sin^2\varphi_i\end{array}\right)\end{array}\right\}. \quad (24)$$

One can see that additional terms in the squared effective EO coefficient appear in case of the isotropic types of AO interaction. These terms, $\chi$ and $\chi^2$, involve the ellipticity of the optical eigenwaves. On the other hand, the relations for $p_{eff}^2$ in case of the anisotropic AO interactions contain only the quadratic terms $\chi^2$. In the limit of zero $\chi$, Equations (16)–(24) reduce to Equations (5)–(13). Notice that the ellipticity $\chi$ can be either positive or negative, so that its sign must be properly accounted for in the effective EO coefficients derived for the isotropic diffraction. Moreover, single-domain lead germanate crystals can be either levorotary or dextrorotary, since their domains with the opposite signs of spontaneous polarization are enantiomorphous. In this work we choose a dextrorotary single-domain crystal. Then the extraordinary optical beam is right-handed while the ordinary beam is left-handed. In this case one has to reverse the sign of the term proportional to $\chi$ in Equations (16), (18) and (20) [21]. Note also that the relations obtained in the work [9] cannot be compared with formulae (16)−(24). This is because the effective EECs in Ref. [9] have been derived for the circularly polarized waves obtained via decomposing the elliptical polarization, while in the present work we deal with elliptically polarized optical waves. Consideration of the influence of piezoelectric effect on the AO figure of merit of lead germanate crystals was not the aim of our present work. Nonetheless, we stress that accounting for piezoelectric contribution into the AW velocities would not lead to the changes in the effective EO coefficients (Eqs. (5)-(13) and (16)-(24)). As a result, this improvement of our consideration would not change the conclusions of our work. However, it would result in some deformation of the angular dependences of the AO figure of merit.

We have verified our theoretical approach using $TeO_2$ as an example. It is known that here the AO figure of merit $1200\times10^{-15}$ $s^3/kg$ is reached under the conditions of anisotropic diffraction when an almost circularly polarized incident optical wave ( $\chi = 0.7$ ) propagates at the angle ~ 1 deg with respect to the optic axis [11]. According to our classification, this corresponds to the type IX of AO interaction. Let the AW $QT_2$ with the frequency 22 MHz propagate along bisector of the *X* and *Y* axes inside the *XY* plane. This wave is polarized in the *XY* plane. The interaction plane $X'Z$ is rotated by 45 deg around the *Z* axis. Figure 2 demonstrates dependences of the effective EO coefficient and the AO figure of merit on the angle $\theta$ between the AW vector and the $X'$ axis. When the dependences are calculated with considering the optical activity, they reveal narrow peaks that correspond to the cases of propagation of AWs close to the $X'$ axis. Then the wavevectors of optical waves are almost perpendicular to the wavevector of AW. They are parallel to the *Z* axis.

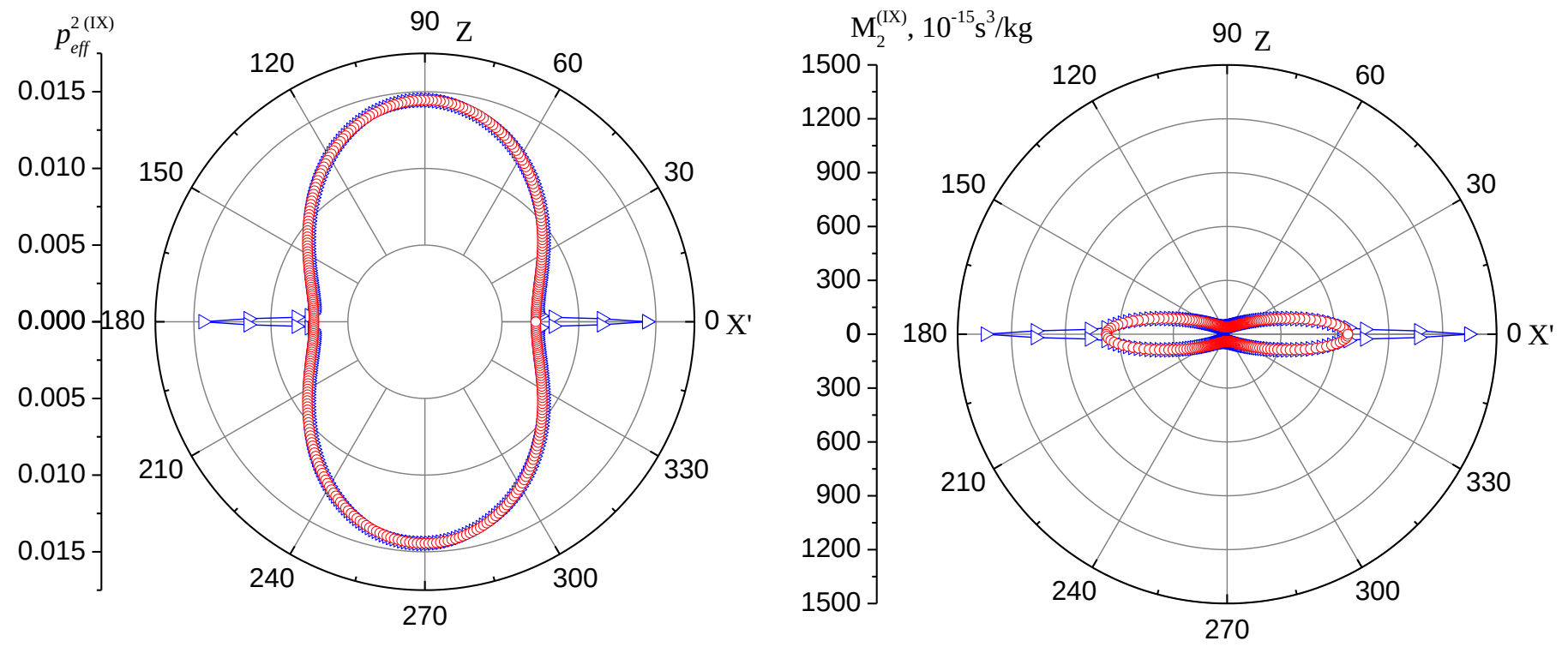

(a) (b)

Figure 2. Dependences of squared effective EO coefficient (a) and AO figure of merit (b) on the angle $\theta$, i.e. on the propagation direction of the AW $QT_2$ in the $X'Z$ plane of $TeO_2$ crystals. Consideration of optical activity corresponds to triangles and its neglect to circles.

It is seen from Figure 2 that accounting for the optical activity effect increases the AO figure of merit from $673\times10^{-15}s^3/kg$ up to $1345\times10^{-15}s^3/kg$. The half-width of the maximum caused by optical activity is about 2.4 deg. The AO figure of merit derived under condition of interaction of the linearly polarized optical waves agrees well with the value obtained in the work [11]. A higher AO figure of merit caused by optical activity also agrees with the experimental data obtained in Ref. [11].

## 3. Results and discussion

Let us deal with the type I of isotropic AO interaction between the optical waves and the QL AW in lead germanate. Then accounting for the ellipticity of optical eigenwaves leads to enhancement of both the effective EO coefficient and the AO figure of merit (Figure 3(a, b)). The half-width of the maximum in AO figure of merit is about 1.6 deg. The AO figure of merit increases more than five times (from $6.8\times10^{-15}s^3/kg$ to $37.9\times10^{-15}s^3/kg$ − see Figure 3(b)). The reason for this increase in the AO efficiency under condition when the ellipticity of eigenwaves is taken into account consists in changing effective EO coefficient. Namely, additional EO tensor components make their contributions into the effective EO coefficient. Primarily, these components are $p_{11}$, $p_{33}$, $p_{13}$ and $p_{31}$, which are two orders of magnitude higher than the coefficients like $p_{15}$, $p_{16}$, $p_{14}$, $p_{51}$, $p_{41}$, $p_{45}$ and $p_{25}$ (see Equations (5) and (16)).

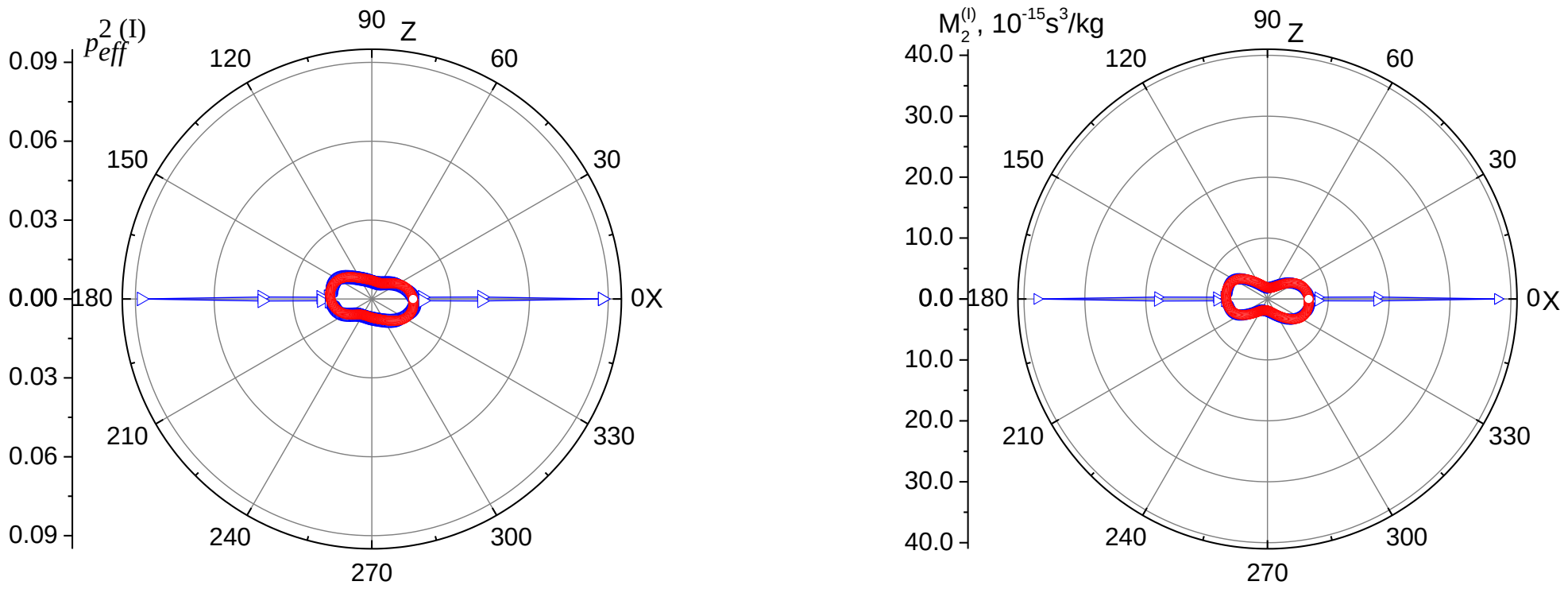

(a) (b)

Figure 3. Dependences of squared effective EO coefficient (a) and AO figure of merit (b) on the angle $\theta$ calculated for $Pb_5Ge_3O_{11}$ crystals when taking into account (triangles) or neglecting (circles) the optical activity: the type I of AO diffraction.

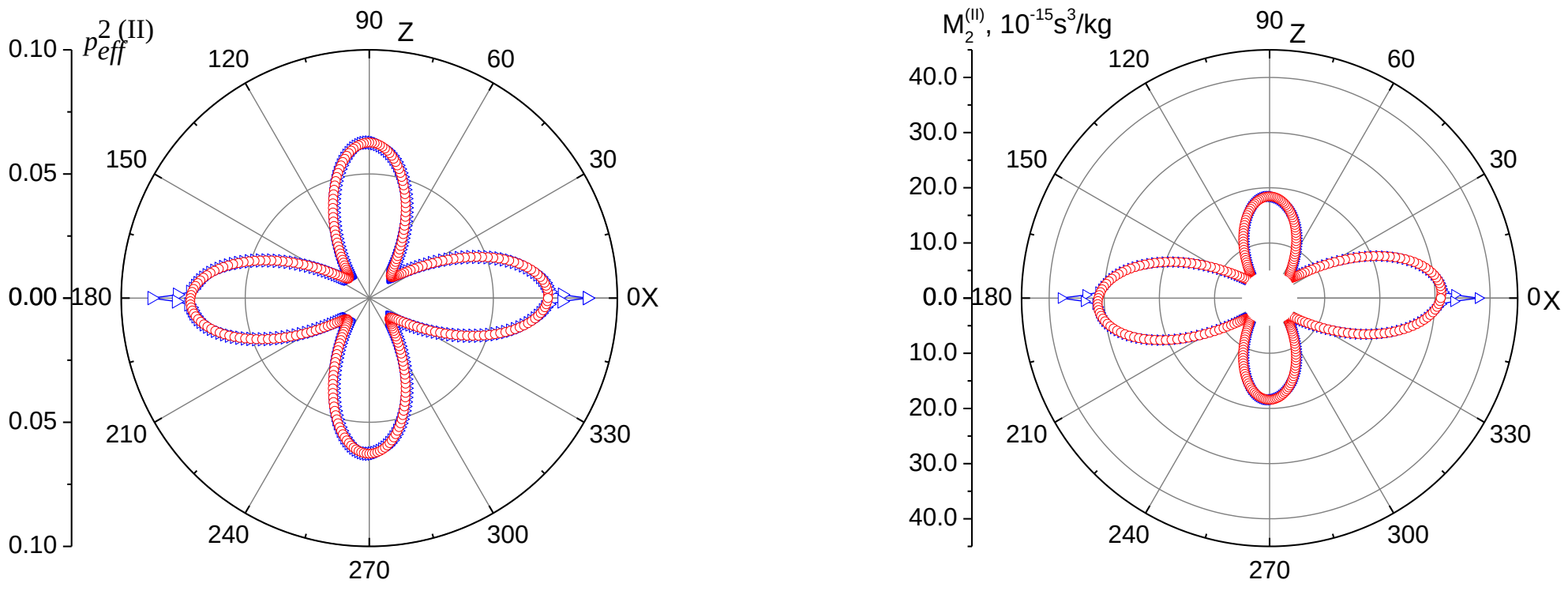

(a) (b)

Figure 4. Dependences of squared effective EO coefficient (a) and AO figure of merit (b) on the angle $\theta$ calculated for $Pb_5Ge_3O_{11}$ crystals when taking into account (triangles) or neglecting (circles) the optical activity: the type II of AO diffraction.

An increase in the effective EO coefficient and the AO figure of merit occurring in a narrow angular region (the half-width of the maximum is equal ~1.6 deg) around the optic axis is also observed for the type II of isotropic AO interaction with the QL AW (Figure 4). The AO figure of merit increases from $31.1\times10^{-15}$ $s^3$/kg to $37.9\times10^{-15}$ $s^3$/kg. This increase is relatively small because the additional contribution into the effective EO coefficient arises only from a small number of EO components, each of which is high enough ($p_{31}$ and $p_{33}$ − see Equations (6) and (17)). The rest of the tensor components with large magnitudes have already been included in the effective EO coefficient in the case of interaction of linearly polarized optical waves (see Equation (6)). Thus, accounting for the optical activity does not affect significantly the AO figure of merit at the type II of AO interactions.

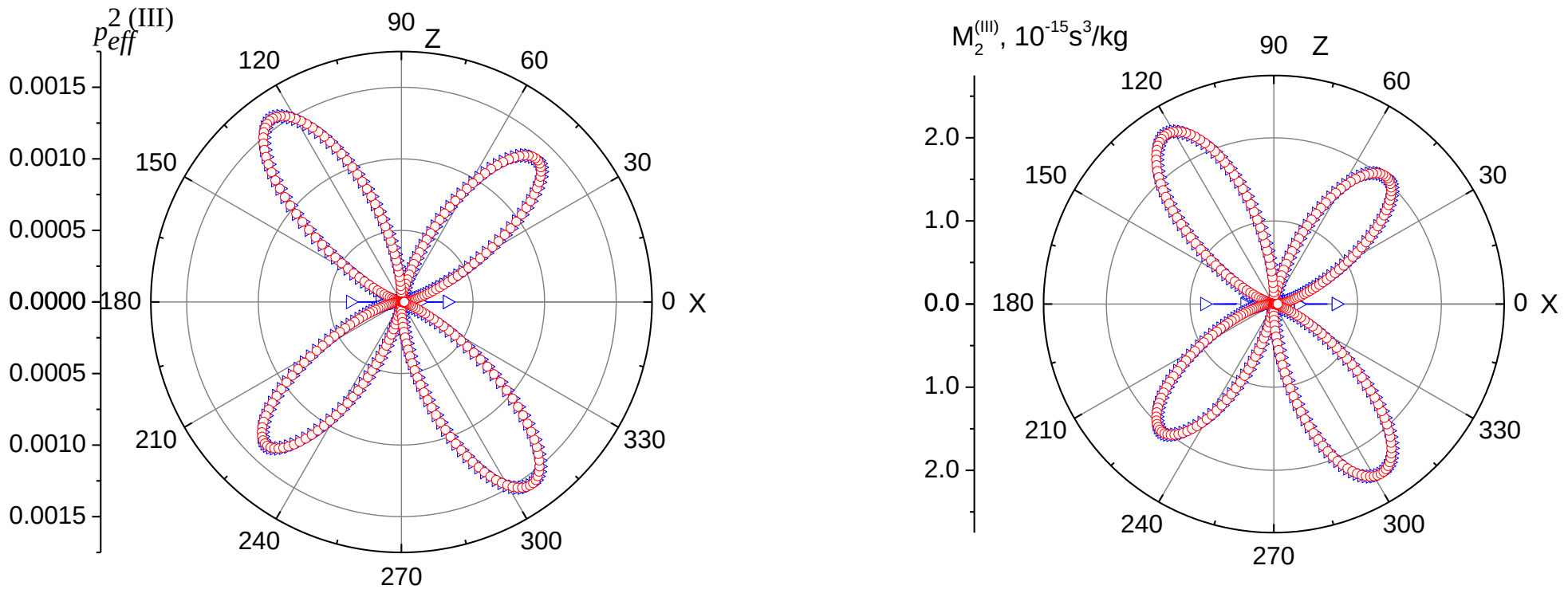

(a) (b)

Figure 5. Dependences of squared effective EO coefficient (a) and AO figure of merit (b) on the angle $\theta$ calculated for $Pb_5Ge_3O_{11}$ crystals when taking into account (triangles) or neglecting (circles) the optical activity: the type III of AO diffraction.

Now we analyze the type III of isotropic AO interaction with the AW $QT_1$. Then the AO figure of merit and the squared effective EO coefficient have respectively the orders of magnitude ~ $10^{-17}$ $s^3/kg$ and $10^{-5}$ whenever the linearly polarized optical waves propagate close to the optic axis (Figure 5). After the ellipticity of optical eigenwaves is accounted for, some additional EO tensor components play a role in AO interactions. Among these components, $p_{11}$, $p_{31}$, $p_{13}$ and $p_{33}$ have large enough magnitudes (cf. Equations (7) and (18)). Their contribution into the AO figure of merit increases this parameter more than order of magnitude (up to $0.74\times10^{-15}$ $s^3/kg$ − see Figure 5). However, the AO figure of merit still remains rather small.

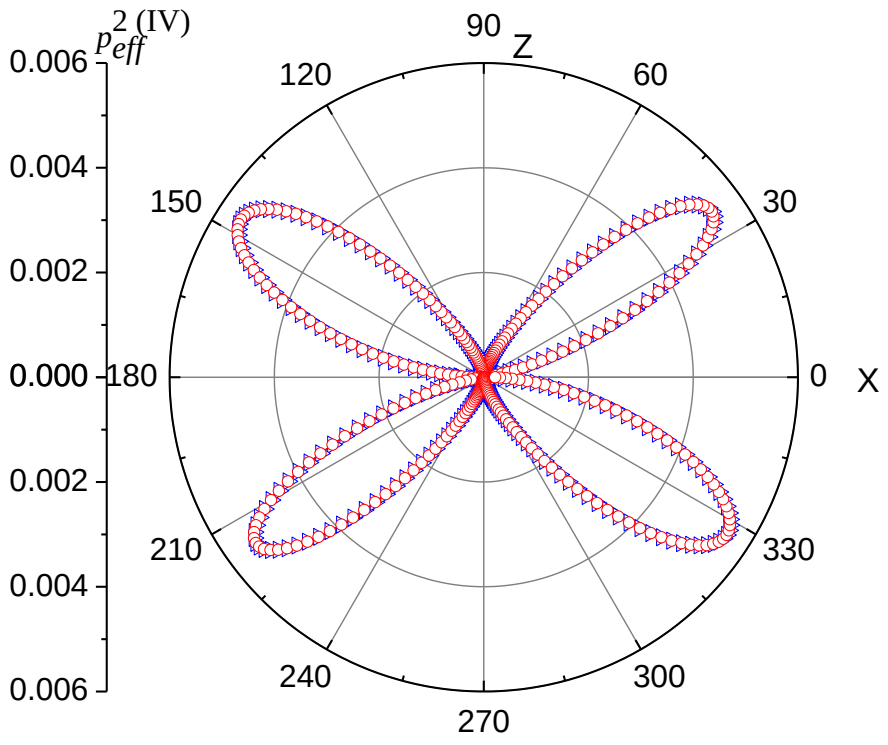

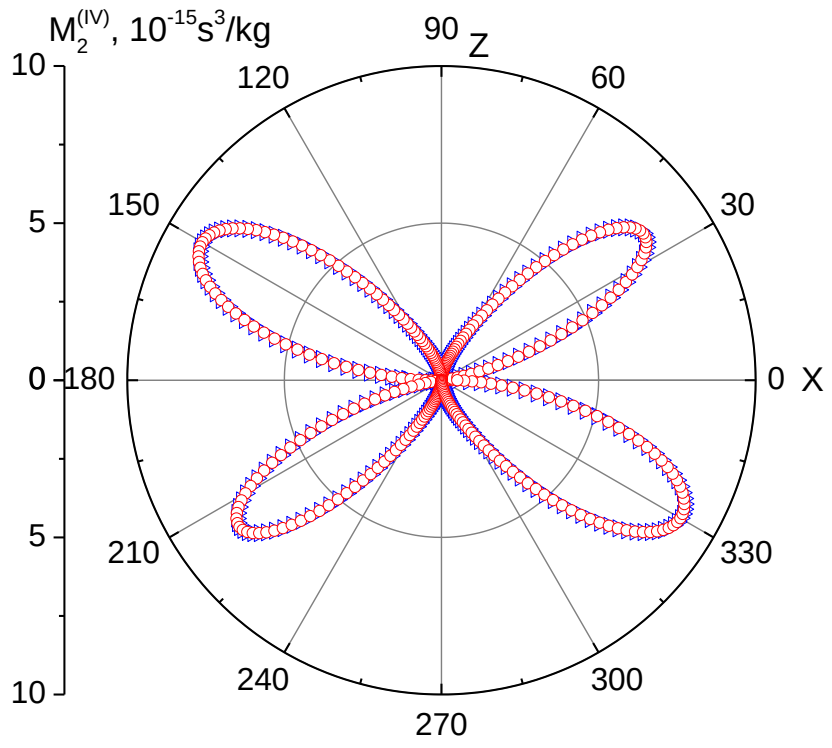

(a) (b)

Figure 6. Dependences of squared effective EO coefficient (a) and AO figure of merit (b) on the angle $\theta$ calculated for $Pb_5Ge_3O_{11}$ crystals when taking into account (triangles) or neglecting (circles) the optical activity: the type IV of AO diffraction.

For the type IV of AO interactions, consideration of the optical rotation makes only minor effect in the effective EO coefficient and the AO figure of merit (Figure 6). To be more exact, a double increase of the AO figure of merit is observed up to the value $0.51\times10^{-15}$ $s^3/kg$. The half-width of the maxima of AO figure of merit, observed at the III and IV types of AO interactions is about 1–2 deg.

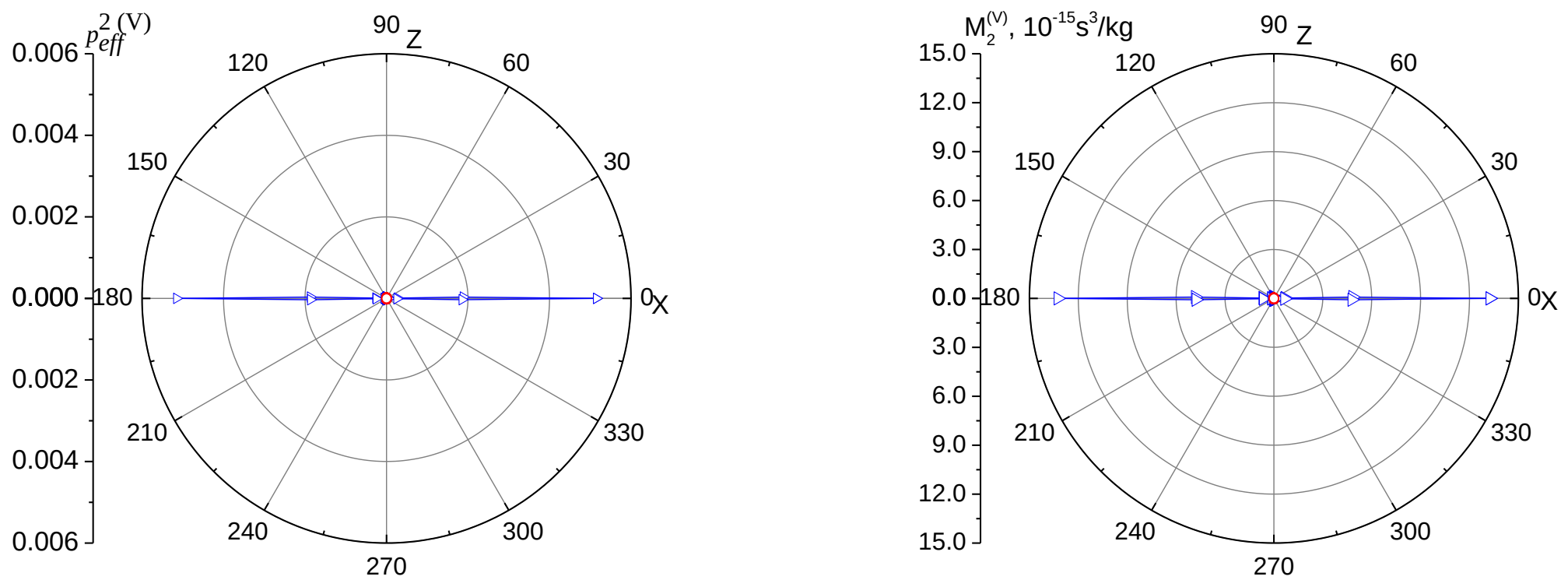

(a) (b)

Figure 7. Dependences of squared effective EO coefficient (a) and AO figure of merit (b) on the angle $\theta$ calculated for $Pb_5Ge_3O_{11}$ crystals when taking into account (triangles) or neglecting (circles) the optical activity: the type V of AO diffraction.

The type V of isotropic AO interaction with the AW $PT_2$ is characterized by a sharp increase in the effective EO coefficient and the AO figure of merit (Figure 7). The AO figure of merit increases from zero up to $13.3\times10^{-15}$ s$^3$/kg. Again, the main role in the enhancement of AO efficiency is played by large enough additional EO components (see Equations (9) and (20)).

The last type VI of isotropic AO interactions with the AW $PT_2$ is also characterized by a notable increase in the effective EO coefficient and the AO efficiency, which is obtained after taking the optical rotation into account (Figure 8). In fact, the AO figure of merit increases from zero up to $13.3\times10^{-15}$ s$^3$/kg. Such enhancement in the AO efficiency is due to contributions of the EO components $p_{44}$ and $p_{66}$, which become involved in the AO interactions owing to the optical activity (see Equations (10) and (21)). The half-width of the maxima of AO figure of merit, observed at the V and VI types of AO interactions is about 4 deg.

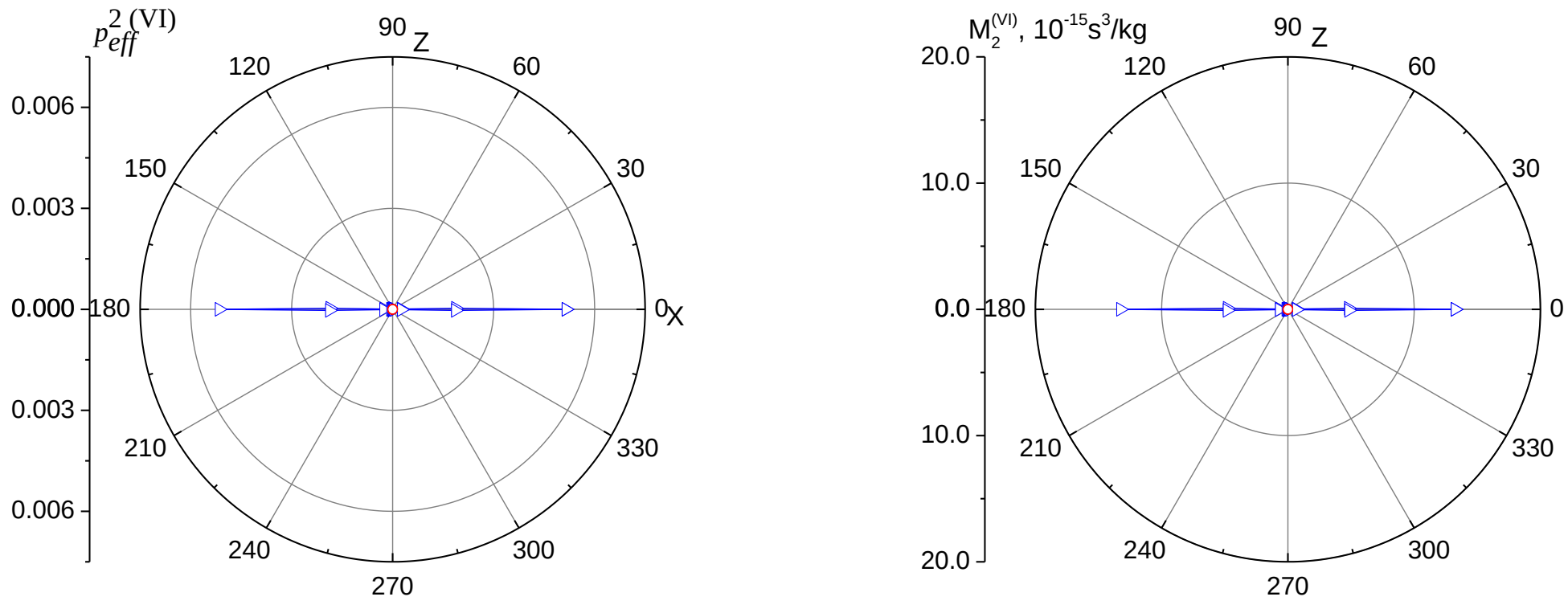

(a) (b)

Figure 8. Dependences of squared effective EO coefficient (a) and AO figure of merit (b) on the angle $\theta$ calculated for $Pb_5Ge_3O_{11}$ crystals when taking into account (triangles) or neglecting (circles) the optical activity: the type VI of AO diffraction.

Now let us consider the anisotropic AO diffraction. Here enhancement of the AO figure of merit due to the optical activity is peculiar for the diffractions of all types (see Figures 9-11). The largest effective EO coefficient and AO figure of merit are peculiar for the type IX of diffraction by the AW $PT_2$ (Figure 11). Then the AO figure of merit increases from $12.5\times10^{-15}$ $s^3/kg$ up to $26.5\times10^{-15} s^3/kg$. The angular half-width of the corresponding peak of AO figure of merit is approximately the same as that observed at the isotropic diffraction and for types VII and VIII of anisotropic diffraction, i.e. its width is about 2 deg around the optic axis. Outside of the angular region ±5 deg, the AO figures of merit calculated in the alternative cases of taking into account or neglecting the ellipticity of optical eigenwaves become nearly the same.

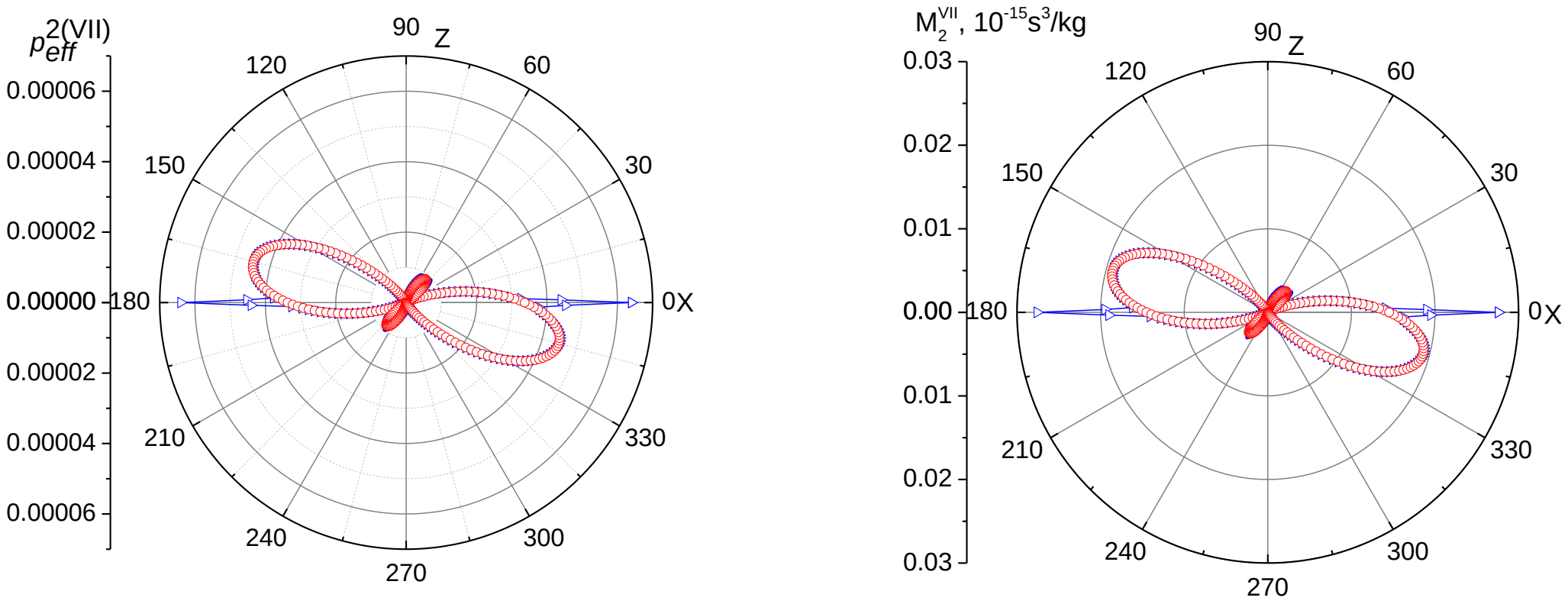

(a) (b)

Figure 9. Dependences of squared effective EO coefficient (a) and AO figure of merit (b) on the angle $\theta$ calculated for $Pb_5Ge_3O_{11}$ crystals when taking into account (triangles) or neglecting (circles) the optical activity: the type VII of AO diffraction.

It is worthwhile that enhancement of the AO figure of merit occurring for all the types of AO interactions is imposed by contributions of additional EO tensor components. These components become involved in the effective EO coefficient due to nonzero ellipticity of the optical eigenwaves, which approaches unity near the optic axis.

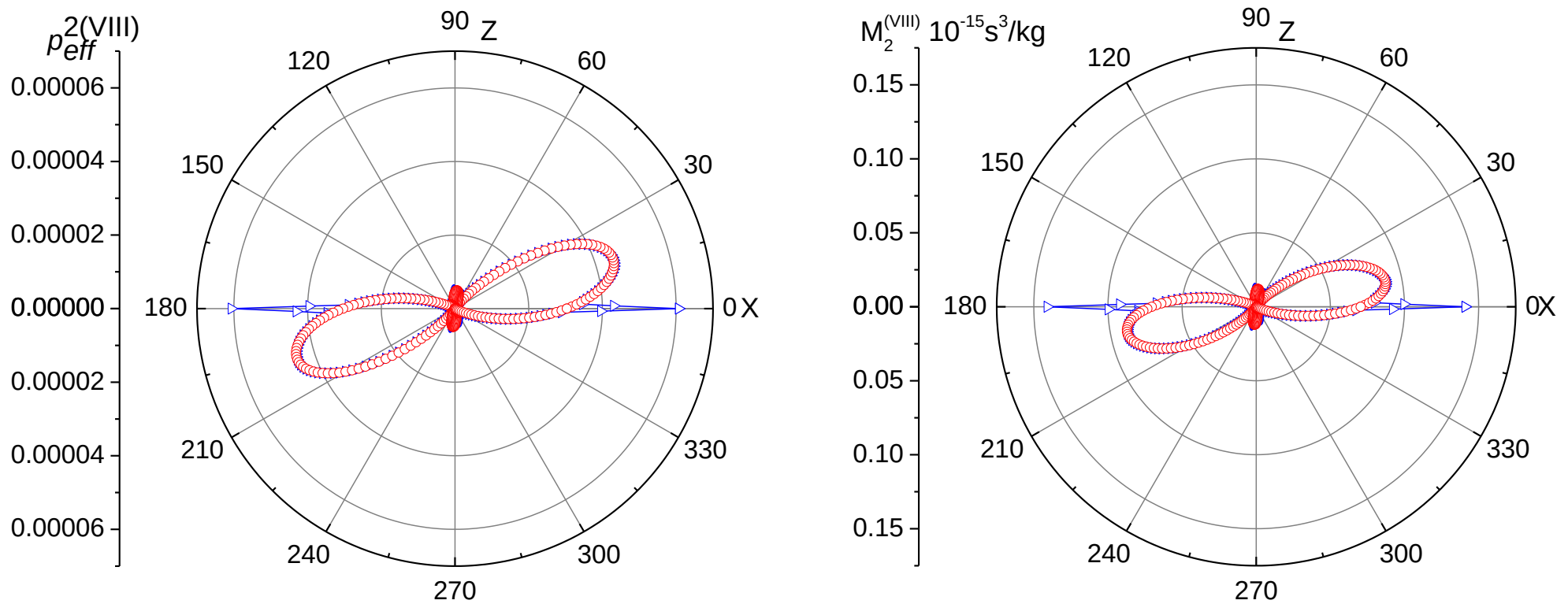

(a) (b)

Figure 10. Dependences of squared effective EO coefficient (a) and AO figure of merit (b) on the angle $\theta$ calculated for $Pb_5Ge_3O_{11}$ crystals when taking into account (triangles) or neglecting (circles) the optical activity: the type VIII of AO diffraction.

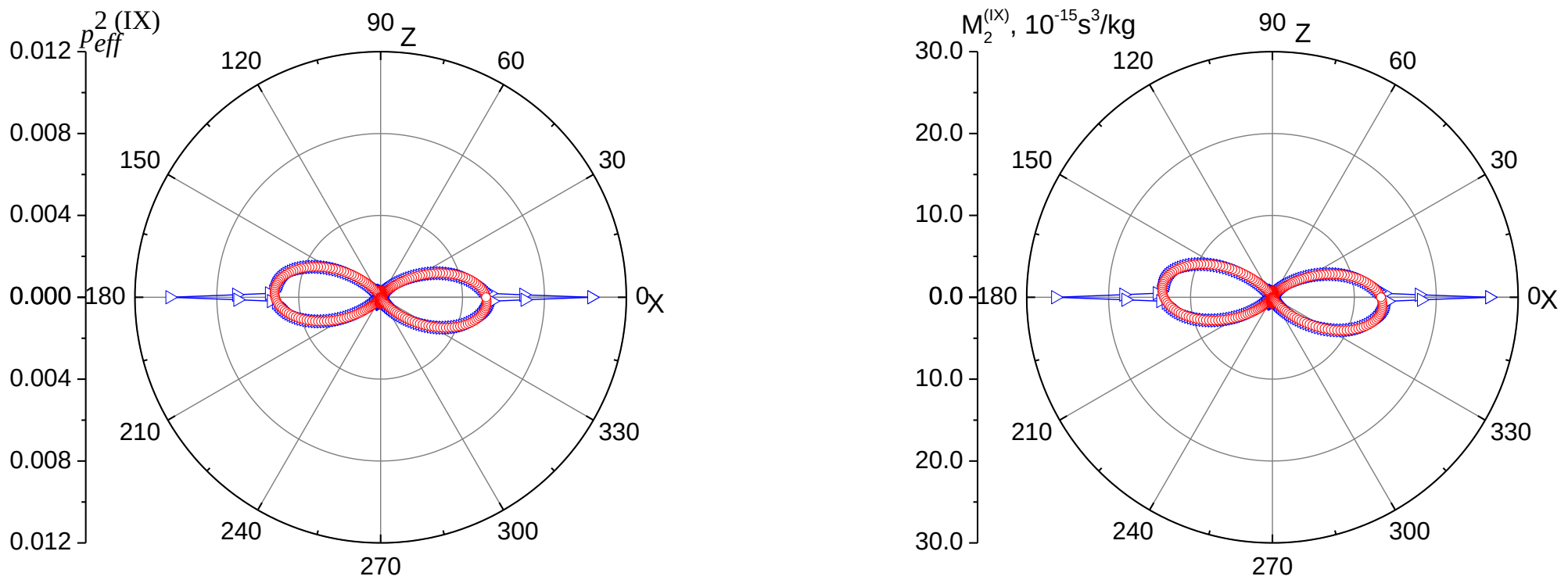

(a) (b)

Figure 11. Dependences of squared effective EO coefficient (a) and AO figure of merit (b) on the angle $\theta$ calculated for $Pb_5Ge_3O_{11}$ crystals when taking into account (triangles) or neglecting (circles) the optical activity: the type IX of AO diffraction.

Finally, we have to note that the optical eigenwaves can acquire elliptical (or circular) polarization not only in non-centrosymmetric crystals possessing the natural optical activity. A similar effect can also be induced by external magnetic (or electric) field in centrosymmetric crystals. In the latter cases, one can operate the efficiency of AO diffraction owing to Faraday (or electrogyration) effect.

## 4. Conclusions

In the present work, we have shown that the existence of optical activity enhances significantly the AO figure of merit. The effect arises due to nonzero ellipticity of the interacting optical eigenwaves. More specifically, the above enhancement takes place because additional EO tensor components become superimposed in the effective EO coefficient. The latter just occurs because the ellipticity of the optical eigenwaves approaches unity in the vicinity of optic axis.

Using the $Pb_5Ge_3O_{11}$ crystals as an example, we have demonstrated that some (larger or smaller) degree of enhancement of the efficiency of AO diffraction is always typical for all the types of isotropic and anisotropic AO interactions, whenever the incident optical wave propagates close to the optic axis. Note also that the AO efficiency can be enhanced not only in the non-centrosymmetric chiral crystals revealing the natural optical activity. Centrosymmetric materials can also manifest induced optical activity which is due to either Faraday effect or electrogyration. To achieve these situations, magnetic or electric field must be applied along the optic axis. This leads to practical possibilities for operating the efficiency of AO diffraction with the external electric and magnetic fields.

We have found that, in a particular case of AO diffraction occurring in the interaction plane *XZ* of $Pb_5Ge_3O_{11}$ crystals, the maximal enhancement of the AO figure of merit takes place at the types V and VI of isotropic AO interactions and at the type IX of anisotropic interaction with the AW $PT_2$. At the type VI of diffraction, the AO figure of merit increases from zero up to $13.3 \times 10^{-15}$ $s^3/kg$, while the AO figure of merit at the type IX of anisotropic diffraction increases almost twice, from $12.5\times10^{-15}$ $s^3/kg$ up to $26.5\times10^{-15}$ $s^3/kg$. Nonetheless, the maximal value of the AO efficiency in the *XZ* interaction plane is reached at the types I and II of isotropic AO interactions. In these cases the AO figure of merit increases from $6.8\times10^{-15}$ up to $37.9\times10^{-15}$ $s^3/kg$ and from $31.1\times10^{-15}$ to $37.9\times10^{-15}$, respectively due to a nonzero ellipticity of optical eigenwaves.

**Acknowledgments**

The authors acknowledge financial support of the present study from the Ministry of Education and Science of Ukraine under the Projects #0121U109804 and #0120U102031.

**Declaration of Interest Statement**

We declare that we have no financial and personal relationships with other people or organizations that can inappropriately influence our work, there is no professional or other personal interest of any nature or kind in any product, service and/or company that could be construed as influencing the position presented in, or the review of, the manuscript entitled.